\documentclass{article}

\usepackage{PRIMEarxiv}
\usepackage{algpseudocode}
\usepackage{amsmath}
\usepackage{subcaption}
\usepackage[numbers]{natbib}
\usepackage{doi}
\usepackage[utf8]{inputenc}
\usepackage[T1]{fontenc}
\usepackage{hyperref}
\usepackage{url}
\usepackage{booktabs}
\usepackage{amsfonts}
\usepackage{nicefrac}
\usepackage{microtype}
\usepackage{fancyhdr}
\usepackage{graphicx}
\usepackage{xspace}
\usepackage{xcolor}
\usepackage{amssymb}
\usepackage{array}
\usepackage{pifont}
\usepackage{threeparttable}
\usepackage{multirow}
\usepackage{listings}
\usepackage{enumitem}
\usepackage{todonotes}
\usepackage{float}
\usepackage{makecell}
\usepackage{caption}
\usepackage{tcolorbox}
\usepackage{authblk}
\usepackage{fontawesome5}

\newcommand{\model}{TurboGR\xspace}

\graphicspath{{media/}}

\title{TurboGR: An Accelerated Training System for Large-Scale Generative Recommendation}

\author{
     \textbf{Huichao Chai}, {Zhixin Wu}, \textbf{Xuemiao Li}, \textbf{Shiqing Fan}, \textbf{Hengfeng Wang}, \textbf{Maojun Peng}, \textbf{Lu Xu}, \\\vspace{0.1cm}
     \textbf{Yaoyuan Wang}, \textbf{Yibo Jin}, \textbf{Wei Guo}, \textbf{Yongxiang Feng}\textsuperscript{\faEnvelope[regular]}\\\vspace{0.2cm}
    
    \normalsize Huawei
}

\begin{document}
\maketitle
\let\thefootnote\relax\footnotetext{\textsuperscript{\faEnvelope[regular]} Corresponding author.}

\begin{abstract}

Generative recommendation (GR) has emerged as a promising paradigm that replaces fragmented, scenario-specific architectures with unified Transformer-based models, exhibiting scaling-law behavior where recommendation quality improves systematically with increased model capacity and training data. However, deploying GR at scale on Ascend NPUs faces fundamental system-level challenges: (1) jagged variable-length sequences cause severe padding redundancy and vector-bound computation, driving MFU below 10\%; (2) the tight coupling of sparse embedding tables and dense Transformer backbones creates communication bottlenecks that limit distributed linearity below 0.6; and (3) token-level negative sampling in long sequences consumes excessive HBM, rendering large-scale recall training infeasible. These challenges are further exacerbated on Ascend NPUs due to the absence of high-performance implementations for jagged operators and the architectural mismatch between irregular sparse primitives and NPU's dense-computation-optimized design. In this paper, we present \model, an Ascend-affinity training system for generative recommendation that systematically addresses these bottlenecks through three core innovations: (i) Ascend-affinity jagged acceleration, including fusion operators that eliminate padding redundancy (achieving 70\% memory reduction and 2.2$\times$ latency speedup) and dynamic load balancing that reduces inter-device imbalance from 47\% to 2.4\%; (ii) distributed communication optimization, comprising hierarchical sparse parallelism (75.9\% all-to-all latency reduction), semi-asynchronous training with proven convergence guarantees, and fine-grained pipeline orchestration that sustains 94\% NPU utilization; and (iii) negative sampling optimization via asynchronous offloading, jaggedness-aware FP16 quantization, and intra-batch logit sharing that expand the effective negative space without additional embedding lookups. Evaluated on the KuaiRand-27K dataset, \model supports training at up to 0.2B parameters and achieves 54.71\% MFU with near-linear scalability (0.97).

\end{abstract}


\section{Introduction}

Recommendation systems now operate as a core computational module in large-scale digital platforms, delivering personalized experiences to billions of users across tens of millions of items~\cite{wu2023personalized, sarwar2000analysis, zannettou2024analyzing}. Modern industrial-scale recommendation systems model diverse signals---collaborative filtering~\cite{sarwar2001item}, sequential behaviors~\cite{kang2018self}, or semantic representations~\cite{wang2014friendbook}---to improve user satisfaction and platform revenue. However, these systems have long suffered from persistent infrastructural bottlenecks: to adapt to a specific scenario, models are often engineered with substantial architectural complexity, making them difficult to generalize and heavily fragmenting the underlying software stack~\cite{zou2025survey, li2025scenario}. This complexity also complicates hardware-level efficiency optimization, frequently leading to low Model FLOPs Utilization (MFU) in traditional deep learning-based recommendation models (DLRMs)~\cite{lin2022building}.

As AI infrastructure evolves towards large-scale paradigms, recommendation systems are moving beyond the traditional DLRM paradigm toward more unified and scalable modeling frameworks, with generative recommendation (GR)~\cite{wang2023generative} emerging as a representative paradigm. While GR adopts unified Transformer architectures, these are specialized recommendation engines structurally and functionally distinct from general-purpose foundation models. By adopting this unified architecture, GR reduces scenario-specific engineering, significantly improves optimization friendliness, and exhibits scaling-law behavior~\cite{zhang2024wukong}: model quality improves systematically with increased data and model capacity, approximately as $Q(C)\approx Q(C_0)+s\log_{4}\!\left(C/C_0\right)$, where $Q(C)$ denotes the relative LogLoss improvement over a fixed baseline and $s$ is the gain per $4\times$ increase in compute complexity. Since the main architecture of GR models is Transformer-based (e.g., HSTU~\cite{zhai2024actions}), optimization techniques developed for LLM training---such as sparse attention and MoE in DeepSpeed~\cite{rasley2020deepspeed}---can be adapted, making the MFU of GR models fundamentally easier to optimize than traditional DLRMs. This compute-bound scaling, however, necessitates highly efficient underlying infrastructure.

However, current open-sourced repositories and the accompanying optimization strategies are mostly GPU-based. Directly deploying these implementations on Ascend NPUs leads to significant performance degradation, stemming from two primary system-level gaps: first, essential operators such as jagged tensor operations lack high-performance NPU implementations; second, fine-grained and irregular parallel primitives (e.g., sorting, top-k, masking) that are highly optimized on GPUs are less efficient on Ascend NPUs, which are primarily optimized for dense matrix computation. These missing and non-dense operators require systematic algorithmic refactoring to fully utilize NPU compute resources.

To bridge this hardware-software gap, we propose \model, an Ascend-affinity training system for generative recommendation. The core contributions of this work are as follows:

\begin{itemize}
    \item We propose Ascend-affinity jagged acceleration techniques, including jagged fusion operators for attention and RAB that eliminate padding redundancy (achieving 70\% memory reduction and 2.2$\times$ latency speedup), jagged embedding lookup acceleration that reduces forward latency by 6$\times$, and dynamic load balancing that reduces inter-device imbalance from 47\% to 2.4\%.
    \item We design distributed communication optimizations comprising hierarchical sparse parallelism (HSP) that reduces all-to-all latency by 75.9\%, a semi-asynchronous training strategy with proven convergence guarantees, and fine-grained pipeline orchestration that sustains 94\% NPU utilization.
    \item We develop negative sampling optimizations via asynchronous offloading (reducing HBM usage by up to 24.59\%), jaggedness-aware FP16 quantization with negligible accuracy loss, and intra-batch logit sharing that expands the effective negative space without additional embedding lookups.
    \item We release \model as the first open-sourced GR system architected for Ascend NPUs, achieving 54.71\% MFU at 0.2B parameters with near-linear scalability.
\end{itemize}

The remainder of this paper is organized as follows. Section~\ref{sec:1-2} presents the system overview and architecture. Section~\ref{sec:evaluation} describes the system evaluation. Section~\ref{sec:system_optimization} details the core system optimization techniques. Section~\ref{sec:conclusion} concludes the paper.

\section{System Overview and Architecture}
\label{sec:1-2}

\begin{figure}[htbp]
    \centering
    \includegraphics[width=0.9\linewidth]{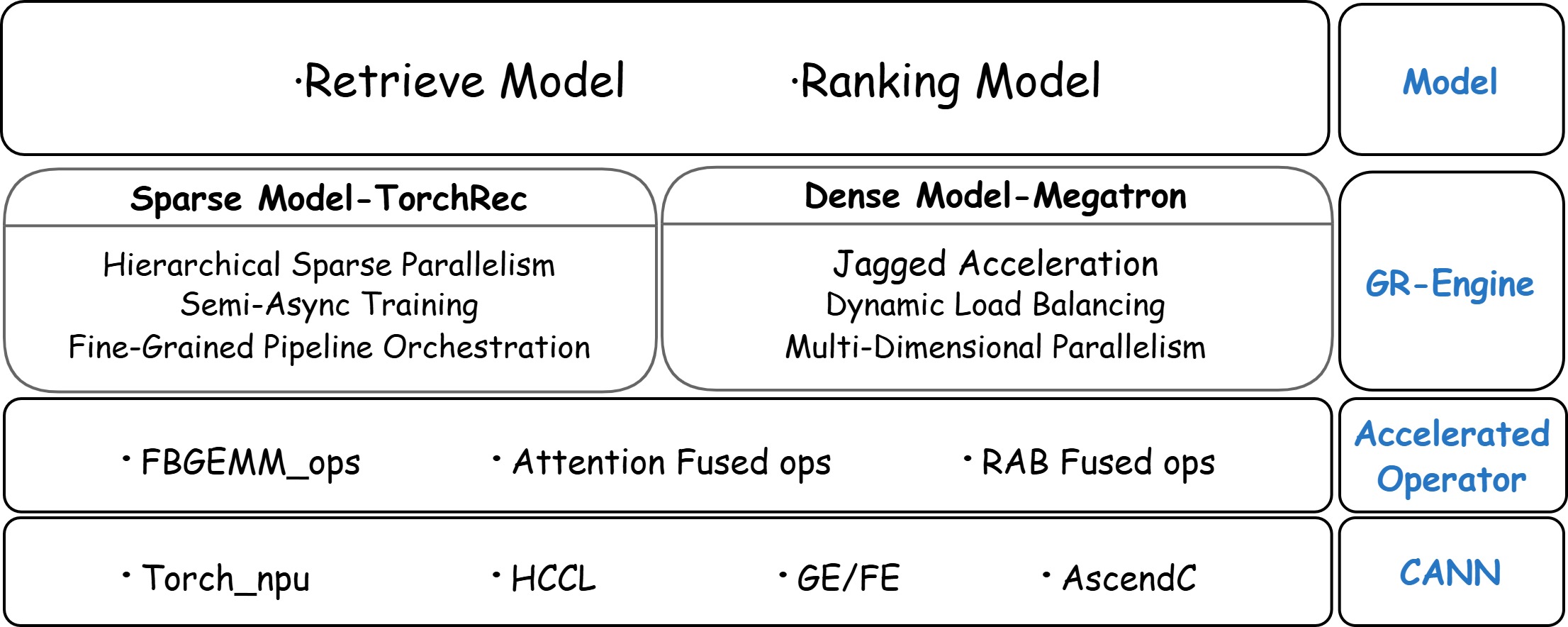}
    \caption{Overall Architectural Design of \model}
    \label{fig:framework}
\end{figure}

The overall architecture of \model is sketched in Figure~\ref{fig:framework}. Designed for large-scale recommendation workloads, the system is organized into modular layers that optimize performance from infrastructure perspectives. The architecture of \model adopts a highly customizable design capable of conducting universal recommendation tasks including retrieval and ranking. To support \model in conducting these recommendation workloads effectively and efficiently, we construct the following underlying layers:
\begin{itemize}
 \item \textbf{GR-Engine}: Serving as the core infrastructural engine, GR-Engine consists of advanced optimization techniques that support large-scale distributed training environments. Specifically, we port established frameworks like TorchRec and Megatron to the Ascend NPU environment and modify their communication backends to be more Ascend-affinity. The detailed introduction of GR-Engine will be demonstrated in Section \ref{sec:system_optimization}.

 \item \textbf{Accelerated Operators}: This layer provides a suite of custom, high-performance operators specifically designed for Ascend NPUs, bypassing the limitations of standard operator libraries. These include jagged fusion operators for attention and relative attention bias (RAB) computation that eliminate padding redundancy, jagged embedding lookup kernels that reduce scalar overhead and improve cache locality, and other NPU-affinity operator adaptations. By deeply exploiting Ascend's hardware characteristics—such as tiling strategies, asynchronous execution, and load balancing across computing elements—these accelerated operators significantly reduce memory footprint and latency, serving as the critical building blocks for efficient GR training. The detailed descriptions are provided in Section \ref{sec:system_optimization}.

 \item \textbf{CANN}: CANN (Compute Architecture for Neural Networks) provides the low-level, full-stack hardware and software structure specifically designed to interface directly with the Ascend platform, handling bottom-level scheduling and memory allocation.
\end{itemize}

\section{System Evaluation}
\label{sec:evaluation}

\subsection{System Deployment Workflow}
\model is architected for seamless industrial deployment and extensive customization. The training workflow proceeds in three stages: (1) \textbf{Data Pipeline Configuration}, where raw interaction logs are preprocessed into sequential sparse inputs with distributed loading optimized to eliminate I/O bottlenecks; (2) \textbf{Model and Topology Instantiation}, where hardware topology (multi-node, multi-NPU) and algorithmic parameters (Transformer variant, QKV dimensions, attention heads) are decoupled and independently configured; and (3) \textbf{Component Extensibility}, where custom fusion operators, attention masking mechanisms, or novel loss functions can be integrated without disrupting the underlying optimized communication backends.

\subsection{Experimental Settings}
\label{sec:4.1}
All experiments are conducted on an Ascend 910B1 64GB NPU cluster (scaling from 32 to 128 NPUs across 4 to 16 nodes) powered by Kunpeng-920 ARM CPUs. We utilize the KuaiRand-27K dataset, applying a 5-core filtering mechanism to eliminate noisy interactions and employing a chronological leave-one-out strategy for the train-test split. The models (HSTU\cite{zhai2024actions}, and FuXi\cite{ye2025fuxi}) are evaluated across multiple scaled variants (tiny, small, medium, large, and long). The models are optimized using AdamW with a learning rate of $4\times10^{-3}$, utilizing TF32 precision and asynchronous embedding updates. The details of experimental settings including the dataset, data preprocessing and parameter setting can be found in Appendix~\ref{appendix:A}.

\subsection{End-to-End System Efficiency}

We evaluated the training efficiency of the system under varying computational workloads. As presented in Table~\ref{tab:Training performance metrics}, the cluster's MFU systematically increases as the model capacity scales up. The gap in MFU across different model sizes demonstrates that larger architectures better exploit NPU tensor core utilization and enhance pipeline efficiency through optimal computation-communication overlapping. Notably, due to its deeply Ascend-affinity architecture, the overall MFU of the FuXi framework consistently outperforms that of HSTU.

\begin{table*}[htbp]
\centering
\small
\caption{Training performance of HSTU and FuXi on KuaiRand-27K (Ascend 910B1 64GB).}
\resizebox{\textwidth}{!}{ 
\begin{tabular}{lcccccccc}
\toprule
\multirow{2}*{\textbf{Model}} &
\textbf{Model Size}   &
\multirow{2}*{\textbf{Seq Lens}} &
\textbf{Comp. Complexity}&
\textbf{Throughput} & 
\textbf{MFU} &
\multirow{2}*{\textbf{Linear Scalability}}  \\
& \textbf{(M)} & &\textbf{(TFLOPs/step)} & \textbf{(sample/s)} & \textbf{(\%)} & \\
\midrule
HSTU-tiny   & 0.17  & 2048 & 0.13 & 3475.17 & 0.43  & 0.38  \\
HSTU-small  & 1.33  & 2048 & 0.59 & 3508.59 & 1.96  & 0.58  \\
HSTU-medium & 10.52 & 2048 & 1.90 & 2896.00 & 8.00  & 0.70  \\
HSTU-large  & 83.97 & 2048 & 4.33 & 1616.83 & 24.74 & 0.93  \\
HSTU-long   & 83.97 & 4096 & 7.15 & 770.79  & 34.08 & 0.97  \\
\midrule
FuXi-tiny   & 0.41   & 2048 & 0.27  & 3441.32 & 0.88  & 0.35 \\
FuXi-small  & 3.18   & 2048 & 1.18  & 3218.31 & 3.78  & 0.50 \\
FuXi-medium & 25.22  & 2048 & 3.13  & 2812.24 & 16.76 & 0.74 \\
FuXi-large  & 201.55 & 2048 & 8.25  & 1156.26 & 39.34 & 0.94 \\
FuXi-long   & 201.55 & 4096 & 11.54 & 574.39  & 54.71 & 0.97 \\
\bottomrule
\end{tabular}
}
\label{tab:Training performance metrics}
\end{table*}

Furthermore, the evaluations of HSTU-long and FuXi-long indicate that longer sequences consistently yield higher MFU, directly attributable to the system's end-to-end redundancy reduction and length-adaptive acceleration capabilities. For both the large and long variants, the cluster linearity strictly exceeds 0.9, as their sustained computational workloads provide sufficient latency hiding to effectively mask distributed communication overheads.

\section{System Optimizations}
\label{sec:system_optimization}
Compared to DLRMs, GR poses significant training challenges due to the tight coupling of large-scale sparse embeddings and deep dense architectures. We formally characterize these challenges as follows.

\paragraph{Challenge 1: Vector-Bound Computation and Jagged Redundancy.}
In GR training, a large number of unique, gather, and hash operators for sparse computation cause vector computation to account for over 80\% of total operations. Additionally, user sequences follow a long-tail distribution, and the resulting jagged tensors contain substantial padding, leading to over 50\% computation redundancy. On Ascend NPUs, where the architecture is optimized for dense matrix computation, these vector-bound and irregular operations yield MFU below 10\% for small-to-medium models.

\paragraph{Challenge 2: Sparse-Dense Communication Bottleneck.}
In GR, both sparse and dense parameters increase according to the scaling law. The embedding table is partitioned across devices via model parallelism, while the dense Transformer backbone requires data-parallel gradient synchronization. This coupling necessitates complex collective communication (e.g., all-to-all for embedding distribution and gradient exchange) and multi-level caching, creating a significant bottleneck that limits distributed linearity below 0.6 in large-scale cluster training.

\paragraph{Challenge 3: Memory-Intensive Negative Sampling.}
Token-level negative sampling in long sequences requires storing a large negative-sample embedding tensor on each NPU. For example, with batch size 8, sequence length 8192, embedding dimension 1024, and 128 negative samples, the negative embedding alone consumes approximately 34\,GB (i.e., $8 \times 8192 \times 1024 \times 128 \times 4$ bytes), resulting in HBM utilization below 50\% and severely limiting scalability under memory constraints.

To address these challenges, we propose GR-Engine, an Ascend-affinity training system that integrates TorchRec for sparse model parallelism and Megatron for dense multi-dimensional parallelism. This system optimizes distributed training across NPU clusters, enabling the efficient integration of hybrid parallelism strategies for both sparse and dense components. The system mitigates the aforementioned bottlenecks through three key innovations: Ascend-affinity Jagged Acceleration (Section~\ref{sec:3.1}), Distributed Communication Optimization (Section~\ref{sec:3.2}), and Negative-Sampling Optimization (Section~\ref{sec:3.3}). 

\subsection{Ascend-affinity Jagged Acceleration}
\label{sec:3.1}

During distributed training of GR models, significant variation in the number of tokens per instance and high sparsity in embeddings pose two fundamental challenges. As a consequence, distributed training suffers from redundant computation, suboptimal memory access efficiency, and severe workload imbalance across workers. To address these issues, we propose an Ascend-affinity acceleration scheme for jagged tensor operations, comprising:

(1) jagged fusion operator for attention and RAB, (2) jagged embedding lookup acceleration, and (3) dynamic jagged load balancing. Together, these techniques enable efficient processing of jagged inputs with variable-length rows and sparse embeddings, leading to improved workload balance, reduced redundant computation, and higher hardware utilization on Ascend NPUs.

\subsubsection{Jagged Fusion Operators for Attention and RAB}\

\begin{figure}
    \centering
    \includegraphics[width=0.8\linewidth]{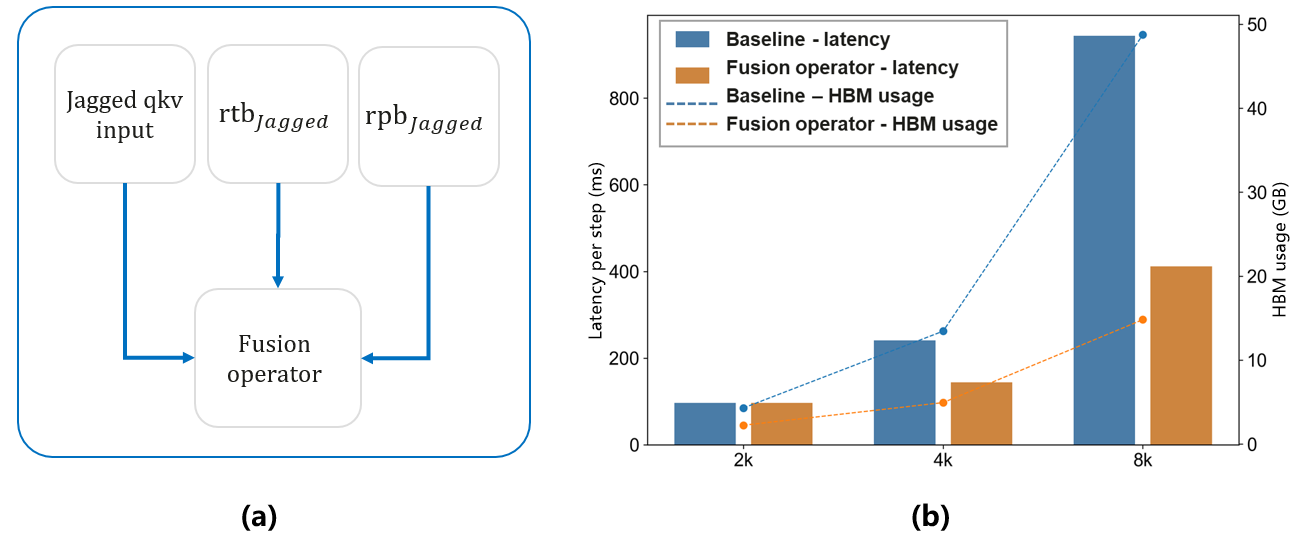}
    \caption{Jagged fusion operators (a) and their efficiency comparison with baseline (b).}
    \label{fig:FusionOps}
\end{figure}

Reducing off-chip memory traffic and maximizing on-chip memory utilization are commonly achieved through fusion operators, which integrate multiple computational steps into a single kernel~\cite{dao2022flashattention,dao2023flashattention}. Thus, to eliminate padding redundancy and accelerate attention and RAB computations, we design Ascend-affinity jagged fusion operators. As illustrated in Figure~\ref{fig:FusionOps}(a), the operators contain jagged qkv input, relative time bias (rtb), and relative position bias (rpb) components. Their implementation leverages input tensor sparsity together with the hardware capabilities of Ascend NPUs. The key performance-enhancing steps are summarized below, with detailed descriptions provided in Appendix~\ref{appendix:rab}.
\begin{enumerate}
    \item \textbf{Eliminating unnecessary conversions.} Mismatches between the jagged format and the model's internal dense representation require costly format conversions (e.g., dense-to-jagged and jagged-to-dense) at the boundaries of each operator. We unify the internal representation so that the entire attention and RAB pipeline operates natively on jagged tensors, removing these conversions entirely and reducing both memory traffic and kernel launch overhead.

    \item \textbf{Tiling strategies and asynchronous execution.} Ascend NPUs process data in tiled chunks to balance compute load and efficiently utilize on-chip cache. We select tiling strategies based on data shapes: rectangular tiling for large attention matrices and square tiling for smaller RAB bias tensors. Furthermore, we exploit the Ascend NPU's capability for asynchronous data copying, overlapping data transfer with computation to hide memory latency.

    \item \textbf{Load balancing across device computing elements.} The RAB backward pass involves a mix of regular (data-parallel) and irregular (index-dependent) operations that naively overload the scalar computing elements. We offload regular computations to vector units and reserve scalar units for irregular operations, requiring only lightweight data packing. This design significantly improves parallelism within each AI Core.

\end{enumerate}

To evaluate the efficiency of the jagged fusion operators, we conduct two groups of controlled experiments: (i) native PyTorch operators as the baseline, and (ii) jagged fusion operators for attention and RAB. The results for the FuXi-long model are illustrated in Figure~\ref{fig:FusionOps}(b). As shown, the synergistic integration of jagged fusion operators leads to substantial performance improvements. Specifically, for 8k sequence length, the end-to-end latency is significantly reduced from 961.21ms to 431.13ms, while the reserved memory footprint decreases from 47.77GB to 14.31GB, corresponding to a 70\% reduction in memory overhead. Overall, the proposed jagged fusion operators significantly reduce the training overhead in both latency and memory usage. 

\subsubsection{Jagged Embedding Lookup Acceleration}\

\begin{figure}
    \centering
    \includegraphics[width=0.6\linewidth]{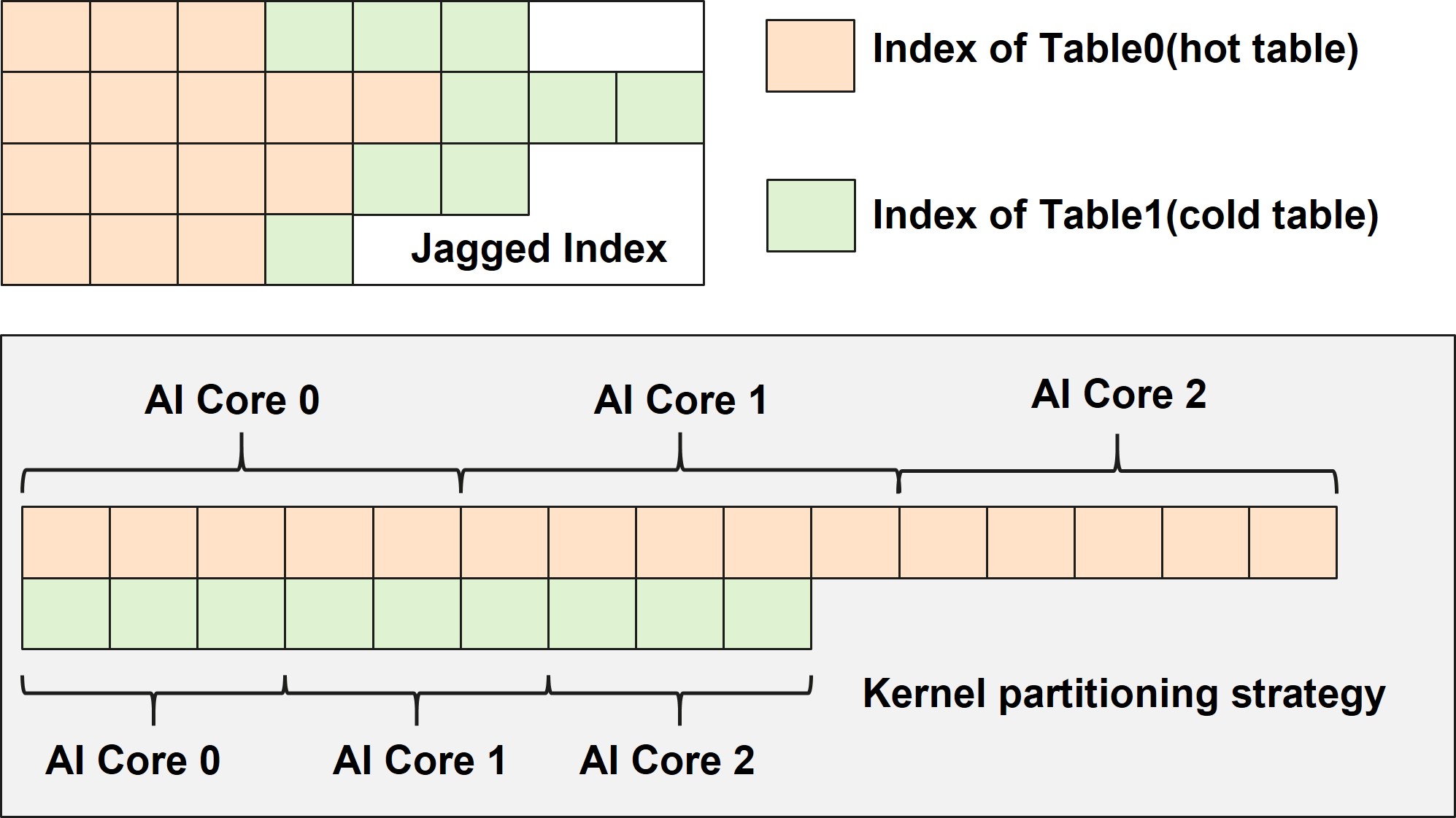}
    \caption{Jagged embedding lookup acceleration.}
    \label{fig:Jagged_Sparse1}
\end{figure}

To address the inefficiency in sparse embedding lookup, we develop a jagged multi-table indexing approach. The overall procedure is described in Figure~\ref{fig:Jagged_Sparse1}. By operating only on valid indices (i.e., ID), it substantially reduces scalar computation and control-flow overhead. 
In addition, batch data are reorganized at the table level and distributed across compute cores, improving cache locality and mitigating load imbalance caused by cold and hot tables. The detailed implementation is described as follows.
\begin{enumerate}
\item[(1)] \textbf{Redundancy Removal and Computation Acceleration.} Padded zeros in variable-length sequences increase the computational overhead in embedding lookup. Therefore, we adopt KeyedJaggedTensor (KJT) data structure implemented in TorchRec~\cite{2022TorchRec}, which operates only on valid indices and avoids unnecessary computation. However, the default KJT implementation still requires additional checks for zero values, introducing redundant scalar operations and conditional branching. To address this issue, we replace the input lengths with explicit jagged indices, thereby fully eliminating padding-induced redundancy.

\item[(2)] \textbf{Kernel Partitioning Strategy.} Na\"ively distributing valid indices across AI Cores leads to single-tiling-block multi-table queries. Embeddings from different tables are typically stored in distant and discrete locations within high bandwidth memory (HBM), which reduces the likelihood of cache hits and consequently increases HBM traffic.

To avoid this, we first restructure batch data at the table level, grouping all data per table across the batch and then distributing each table's data evenly across all AI cores, as is shown in the bottom part of Figure~\ref{fig:Jagged_Sparse1}. This design enables synchronized processing of identical tables across cores, improving L2 cache hit rates and mitigating workload imbalance due to hot and cold table access patterns. 
\end{enumerate}

The experimental results in Table~\ref{tab:Jagged EC} show that, for a batch of ~1 million IDs (with 50.43\% invalid padded zeros), the forward latency of the embedding lookup is reduced by 6 times, while the backward latency is reduced by 4 times. This confirms that jagged embedding lookup acceleration is essential for efficiently supporting sparse embeddings in GR multi-table training.

\begin{table}
    \caption{Jagged embedding lookup latency comparison.}
    \centering
    \resizebox{\textwidth}{!}{ 
    \begin{tabular}{ccccc}
    \toprule
         \multirow{2}{*}{} & 
         \multirow{2}{*}{Total Indices} & 
         \multirow{2}{*}{Padded Zeros} & 
         \multicolumn{2}{c}{Embedding Lookup Latency} \\
         \cmidrule(lr){4-5}
         & & & \multicolumn{1}{c}{Forward (ms)} & \multicolumn{1}{c}{Backward (ms)} \\
         \midrule
         Baseline & \multirow{2}{*}{1064960} & \multirow{2}{*}{537019} & 18 & 36 \\
         Jagged embedding lookup acceleration &  & & 3 & 9 \\
    \bottomrule
    \end{tabular}
    }
    \label{tab:Jagged EC}
\end{table}

\subsubsection{Dynamic Jagged Load Balancing}\

\begin{figure}
    \centering
    \includegraphics[width=0.9\linewidth]{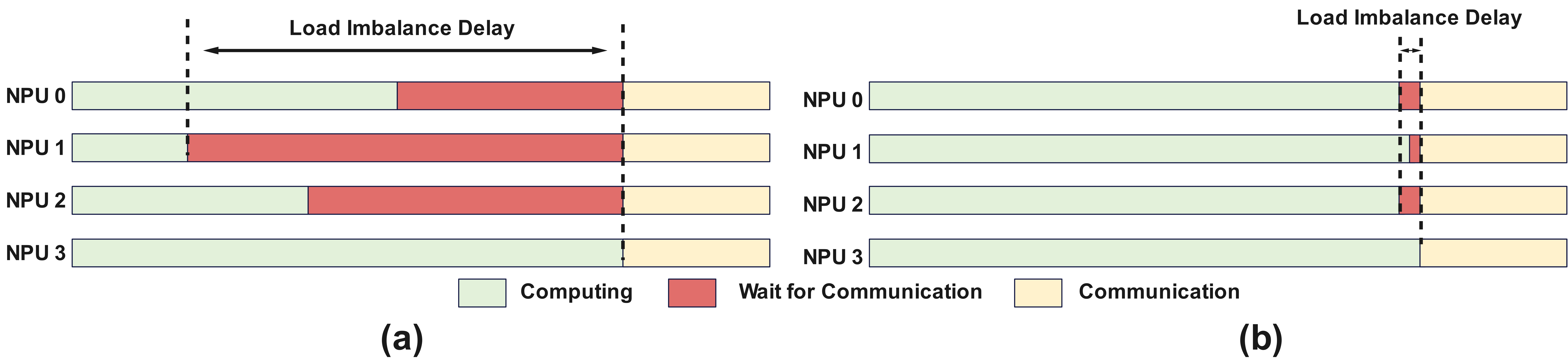}
    \caption{Training timeline of the baseline (a) and the dynamic jagged load balancing (b).}
    \label{fig:Dynamic_Sequence2}
\end{figure}

\begin{figure}
    \centering
    \includegraphics[width=0.9\linewidth]{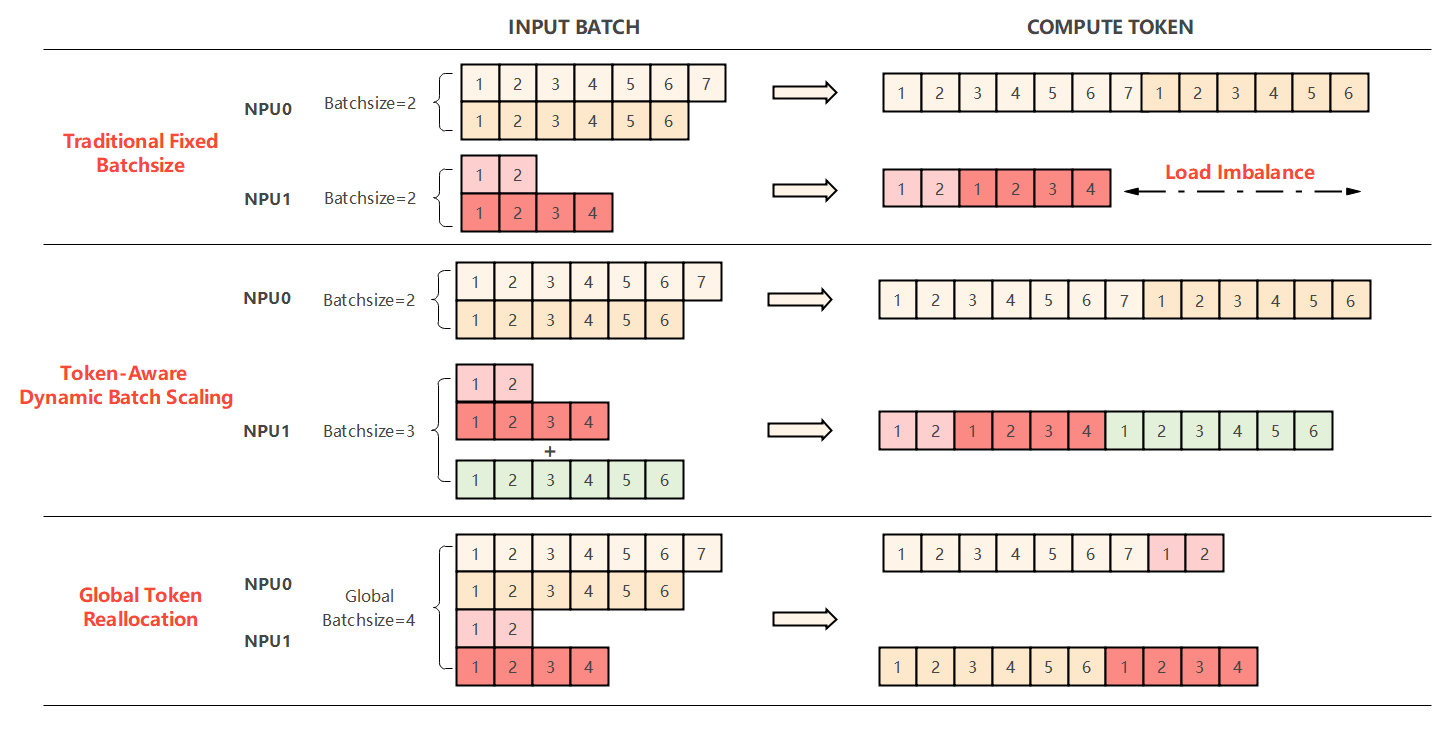}
    \caption{Token-aware dynamic batch scaling and global token reallocation.} 
    \label{fig:DLB scheme}
\end{figure}

In data-parallel GR training, variable-length multi-feature sequences cause computational imbalance across workers, leading to severe synchronization delays at collective communication barriers and reduced training efficiency. For example, Figure~\ref{fig:Dynamic_Sequence2}(a) illustrates a multi-NPU training scheme, where devices with lighter workloads spend nearly half their time idling at synchronization points waiting for heavily loaded devices. To address this computational imbalance, we propose two complementary strategies, as shown in Figure~\ref{fig:DLB scheme}:
\begin{itemize}
\item[\textbullet] \textbf{Token-Aware Dynamic Batch Scaling.} This strategy targets short-sequences, where batch sizes are adjusted based on token counts rather than fixed sample counts. During data loading, each NPU dynamically scales its batch size according to a predefined token threshold derived from the baseline. 
As a result, each NPU processes a comparable number of effective tokens per step, leading to more balanced computational workloads. Since dynamic batch scaling causes the sample count per NPU to vary, we further adopt a sample-count-weighted gradient aggregation strategy to preserve consistent optimization behavior across NPUs.

\item[\textbullet] \textbf{Global Token Reallocation.} This strategy targets long-sequence scenarios with small batch size, where tokens are distributed across devices to balance workloads without breaking sequence integrity. We restructure the data loading process to construct a global batch across all NPUs, sort samples by token count, and assign them to devices using a greedy load-balancing strategy. As a result, each device receives a set of samples with comparable token counts, significantly improving workload balance and training throughput in multi-NPU settings.

\end{itemize}

We conduct ablation studies on the short-sequence Amazon-all dataset to evaluate the token-aware dynamic batch scaling strategy, and on the long-sequence KuaiRand-27K dataset to evaluate the global token reallocation strategy. To ensure a fair comparison, we randomly sample profile data from the same set of 16 NPUs at identical training steps across all experimental settings.

As shown in Table~\ref{tab:Dynamic Sequence Balancing}, on the Amazon-all dataset, 
the maximum token count difference across NPUs is reduced from 623 to 31, while the proportion of load-imbalance-induced communication latency in the end-to-end runtime decreases from 3.55\% to 1.48\%. On the KuaiRand-27K dataset, the maximum token count difference drops drastically from 10,726 to 559, and the corresponding communication latency ratio is reduced from 47.01\% to 2.40\%.

Figure~\ref{fig:Dynamic_Sequence2}(b) provides an intuitive visualization of the optimized training scheme, in comparison with the baseline shown in Figure~\ref{fig:Dynamic_Sequence2}(a). These results demonstrate that the proposed strategy substantially improves synchronization efficiency across devices in distributed training, thereby enhancing overall computational resource utilization.

\begin{table}
    \centering
    \small
    \caption{Performance comparison with and without the load balancing optimization.}
    \resizebox{\textwidth}{!}{ 
    \begin{tabular}{cccccc}
    \toprule
         \multirow{2}{*}{\textbf{Dataset}} &
         \multirow{2}{*}{\textbf{Strategy}} &
         \textbf{Maximum Token} &
         \textbf{Single-step}&
         \textbf{Load Imbalance } &
         \textbf{Load Imbalance }\\
         &&\textbf{Count Difference}&\textbf{Latency(ms)}&\textbf{Delay(ms)}&\textbf{Ratio(\%)}\\
    \midrule
    \multirow{2}{*}{\raisebox{-0.5\height}{Amazon-all}}
      & \makecell{Fixed Batchsize\\Baseline}
      & 623 & 422 ms & 15 ms & 3.55\\
      & \makecell{Token-Aware\\Dynamic Batch Scaling}
      & 31 & 405 ms & 6.5 ms & 1.48\\
    \midrule
    \multirow{2}{*}{\raisebox{-0.5\height}{KuaiRand-27K}}
      & \makecell{Fixed Batchsize\\Baseline}
      & 10726 & 2340 ms & 1100 ms & 47.01\\
      & \makecell{Global Token\\Reallocation}
      & 559 & 1540 ms & 37 ms & 2.40 \\
    \bottomrule
    \end{tabular}
    }
    \label{tab:Dynamic Sequence Balancing}
\end{table}

\subsection{Distributed Communication Optimization}
\label{sec:3.2}

\begin{figure}
    \centering
    \includegraphics[width=0.8\linewidth]{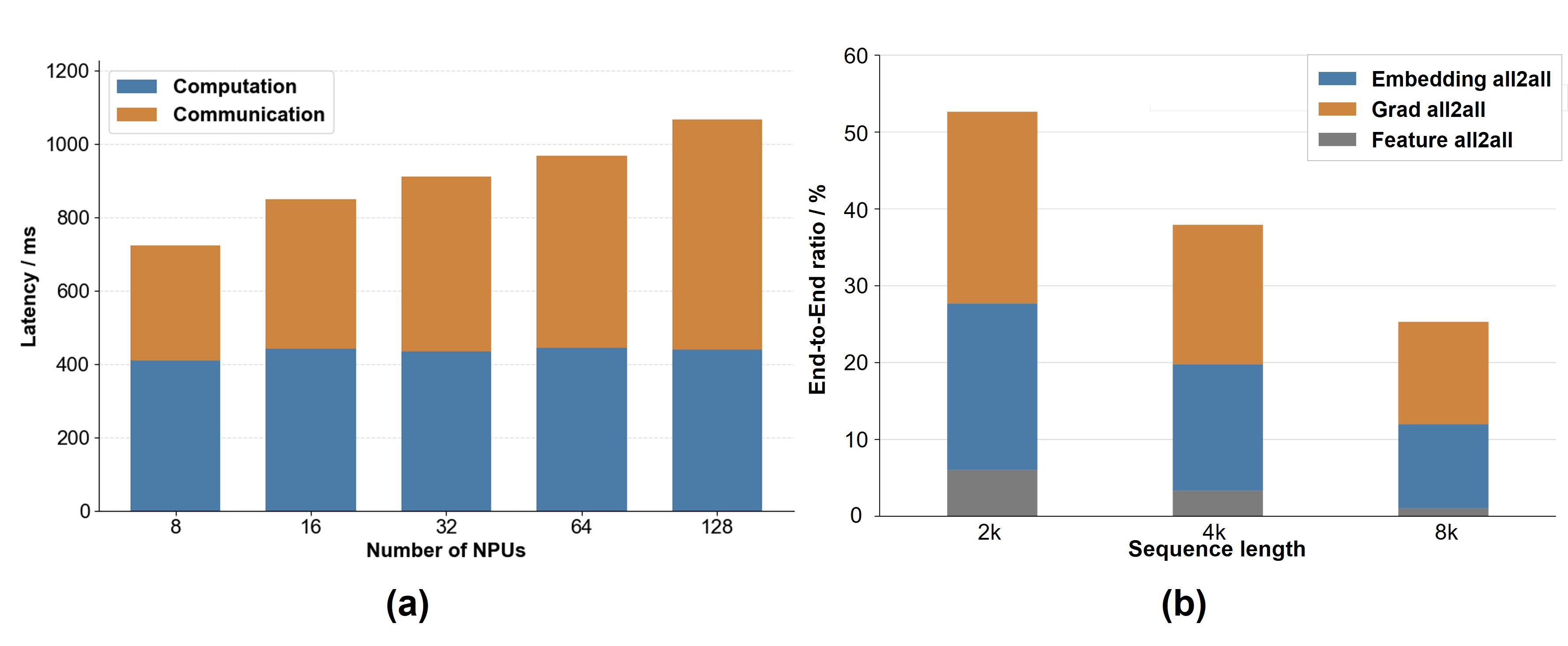}
    \caption{Latency breakdown versus NPU count (a) and sequence length (b).}
    \label{fig:HSP Communication time varies}
\end{figure}

In distributed GR training, scaling to larger clusters improves computational efficiency but simultaneously incurs substantial communication overhead. As shown in Figure~\ref{fig:HSP Communication time varies}(a), distributed communication has emerged as the primary bottleneck constraining overall training throughput. To sustain cluster-level scalability, we introduce a comprehensive set of distributed parallelism and pipeline optimizations. Specifically, (1) to mitigate communication overhead induced by partitioned embeddings, we propose adaptive hierarchical sparse parallelism; (2) to reduce the all-to-all communication overhead, we adopt a semi-asynchronous training strategy; and (3) to minimize NPU idle time, we further design a fine-grained pipeline orchestration that operates on the end-to-end training procedure.

\subsubsection{Adaptive Hierarchical Sparse Parallelism}\
\label{sec:3.2.1}

\begin{figure}
    \centering
    \includegraphics[width=1.0\linewidth]{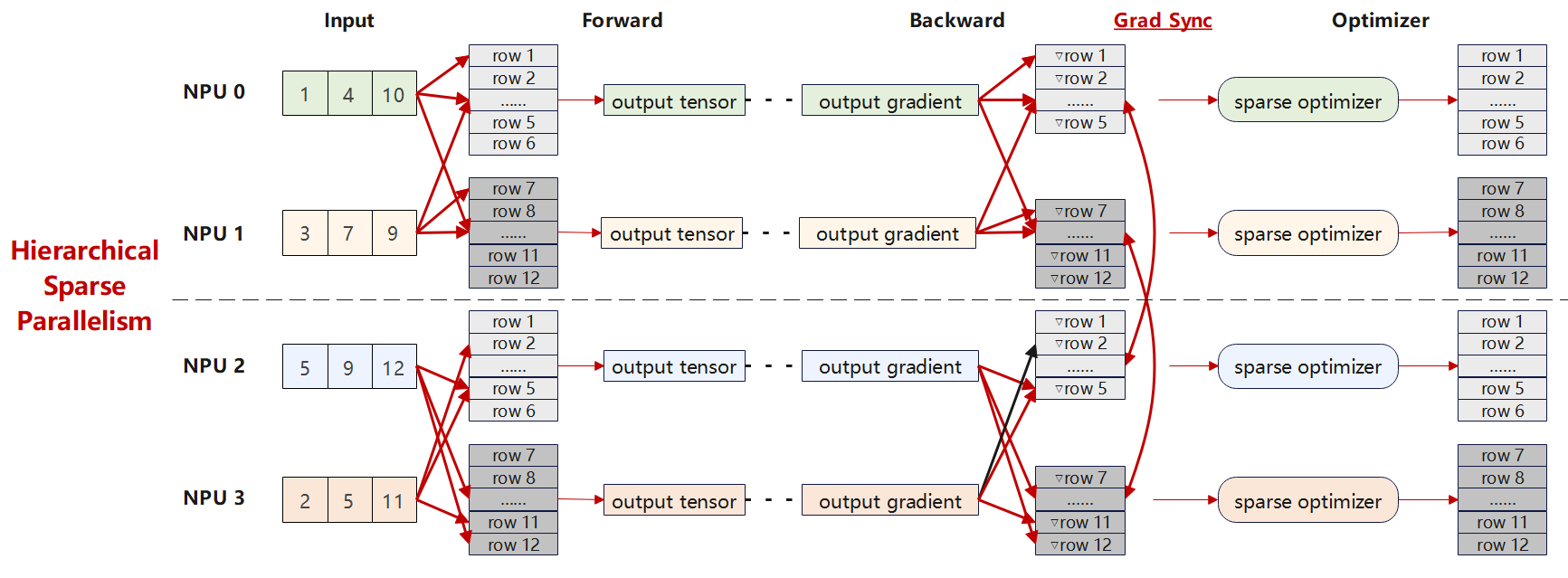}
    \caption{Hierarchical sparse parallelism (HSP).}
    \label{fig:HSP scheme}
\end{figure}

In the standard TorchRec implementation, embedding tables are partitioned across all devices, requiring two-phase global all-to-all collectives to align input features and embedding lookup results~\cite{2022TorchRec}. As cluster scale increases, communication overhead becomes the dominant bottleneck, severely limiting training scalability. To address this issue, we adopt a restructured embedding partitioning and communication architecture, namely hierarchical sparse parallelism (HSP), as illustrated in Figure~\ref{fig:HSP scheme}. Given a distributed setup with $N$ devices, we organize the devices into $M$ parallel groups, each containing $I = N/M$ devices. Each group maintains a full replica of the embedding table, which is internally partitioned via model parallelism. During training, each group performs embedding lookups only for the IDs appearing in its local mini-batch, while all-to-all communication is confined within the group. Consequently, HSP accelerates all-to-all communication by reducing the total number of lookup IDs participating in global collectives and alleviates cross-group communication overhead.

With HSP, we reduce the communication scale from O(N) to O(I) thereby mitigating the communication degradation as the device count grows.

Meanwhile, we apply data parallelism across groups. Unlike existing implementations such as~\cite{zhang2025two}, which modify the ``moment scale'' to compensate for the reduced effective learning rate induced by weight synchronization, HSP avoids this algorithmic complexity by directly reconstructing the backward operator for embedding lookup. Specifically, we developed two strategies to reduce the communication overhead. First, by re-engineering the operator, we introduce an all-reduce communication to achieve gradient synchronization between groups before the update step. It ensures that all parallelism groups perceive an identical aggregate gradient $G_t = \sum_{j=1}^{M}g_{j, t}$. With the AdaGrad optimizer, the state $S$ and parameter $W$ updates for each group $i$ at each step $t$ are given by:
\begin{equation}
\begin{aligned}
    S_{i, t} &= S_{i, t-1} + G_t^2,\\
    W_{t+1} &= W_t - \frac{\eta}{\sqrt{S_{i, t}+\varepsilon}}\cdot G_t.
    \end{aligned}
\end{equation}

Since the initial states $S_{i, 0}$ are identical and all groups receive the uniform aggregate gradient $G_t$, the locally maintained optimizer states (second-order moments) evolve identically across all groups, such that $S_{1,t}=S_{2,t}=\cdots=S_{M,t}$. Consequently, the global optimization trajectory remains mathematically equivalent to standard centralized training, ensuring lossless convergence without the need for manual learning rate scaling. Second, we implement a sparse gradient exchange mechanism that transmits only the indices and values of activated gradient entries, instead of synchronizing the full embedding table. This sparse synchronization is executed on dedicated streams, enabling asynchronous overlap of communication with the dense computation in subsequent layers, and thus preventing synchronization from becoming a new bottleneck.

We analyze the all-to-all communication overhead, total communication latency, and end-to-end training latency of HSP in Table~\ref{tab:HSP}. The results demonstrate the effectiveness of HSP, achieving a 75.9\% reduction in all-to-all communication latency, while the overall communication latency decreases by 39.1\%. Despite introducing additional all-reduce communication and modest computational overhead, HSP still delivers a significant improvement in end-to-end training latency. Overall, HSP improves TorchRec scalability by reconciling hierarchical partitioning with gradient-level synchronization, enabling efficient training at scale.

\begin{table}
    \centering
    \caption{Performance comparison with and without the HSP optimization.}
    \begin{tabular}{cccc}
    \toprule
         \multirow{2}*{\textbf{Strategy}} &
         \textbf{Single-step} & \textbf{all-to-all} & \textbf{Overall Communication}\\
         &\textbf{Latency (ms)}&\textbf{Delay (ms)}&\textbf{Latency (ms)}\\
         \midrule
         Baseline & 1970 & 498 & 613 \\
         HSP & 1790 & 120 & 373 \\
    \bottomrule
    \end{tabular}
    \label{tab:HSP}
\end{table}

\subsubsection{Semi-Async Training Strategy}\
\label{sec:3.2.2}

In TorchRec, embedding tensor parallelism is implemented via all-to-all collectives, consisting of three key phases: sparse feature distribution, embedding vector exchange, and gradient synchronization. These phases incur substantial latency-bound communication, leading to severe under-utilization of computational resources. As shown in Figure~\ref{fig:HSP Communication time varies}(b), sparse communication accounts for 53\% of end-to-end time with a 2k sequence and 25\% with an 8k sequence. Within sparse communication, embedding transfer and gradient synchronization dominate; for the 2k case, they contribute 22\% and 25\% of total time, respectively. Because the standard execution flow is strictly sequential, these embedding and gradient communications cannot be overlapped with dense computation.

\begin{figure}
    \centering
    \includegraphics[width=0.8\linewidth]{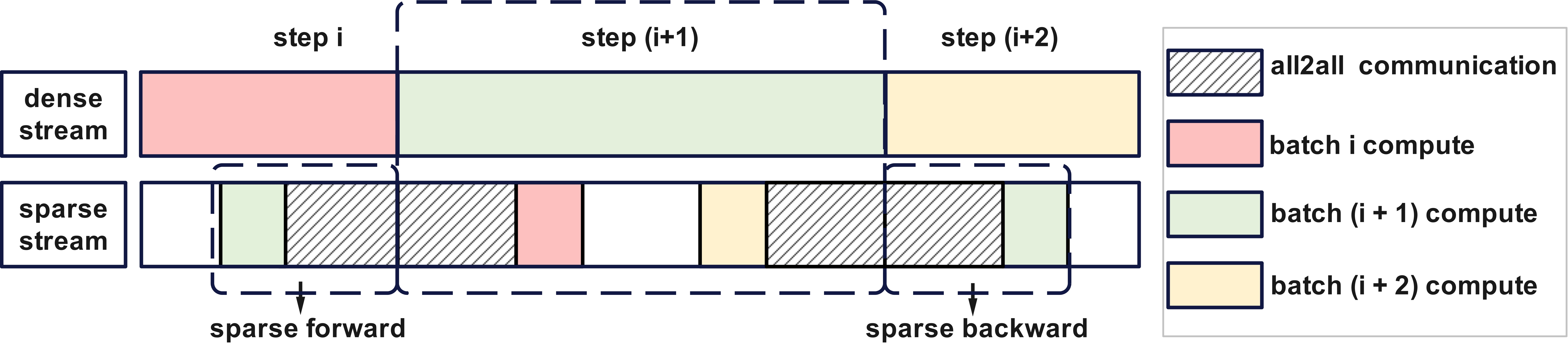}
    \caption{Semi-async training strategy.}
    \label{fig:Semi-Async scheme}
\end{figure}

To tackle the all-to-all bottleneck, we adopt a semi-async training strategy with sparse-asynchronous and dense-synchronous execution. Specifically, sparse asynchrony removes the dependency of the sparse forward pass of batch $(i+1)$ on the sparse backward pass of batch $i$, allowing the sparse forward of batch $(i+1)$ to run ahead of the sparse backward of batch $i$, as illustrated in Figure~\ref{fig:Semi-Async scheme}. This design effectively advances sparse execution by one step across batches, while preserving the original inter-batch dependencies for the dense computation. We postpone the convergence of the strategy to Appendix~\ref{sec:convergence}.

\begin{table}[]
    \centering
    \caption{Performance comparison of the semi-async training strategy.}
    \resizebox{\textwidth}{!}{ 
    \begin{tabular}{lccccccc}
    \toprule
    \multirow{2}*{\textbf{Method}} & \multicolumn{2}{c}{\textbf{Unmasked Sparse Comm.}} & \multirow{2}*{\textbf{HR@10}} & \multirow{2}*{\textbf{HR@200}} & \multirow{2}*{\textbf{HR@2000}} & \multirow{2}*{\textbf{NDCG@10}} & \multirow{2}*{\textbf{NDCG@200}} \\
    \cmidrule(lr){2-3}
    & \textbf{Time (ms)} & \textbf{Ratio (\%)} & & & & & \\
    \midrule
    Baseline    & 459.29 & 24.12 & 0.0295 & 0.1757 & 0.4528 & 0.0151 & 0.0404 \\
    Semi-Async  & 29.37  & 2.19  & 0.0306 & 0.1779 & 0.4516 & 0.0159 & 0.0414 \\
    \bottomrule
    \end{tabular}
    }
    \label{tab:Semi-Async training comm}
\end{table}

We examine the efficiency and recommendation performance of the proposed semi-asynchronous training strategy. As shown in Table~\ref{tab:Semi-Async training comm}, it significantly reduces latency and the unmasked collective communication ratio compared to the baseline, as embedding and gradient communications are almost entirely masked. In addition, recommendation performance is retained, since convergence is theoretically guaranteed. Overall, semi-async training effectively reduces communication overhead while preserving the original recommendation performance.

\subsubsection{Fine-Grained Pipeline Orchestration}\

\begin{figure}
    \centering
    \includegraphics[width=0.6\linewidth]{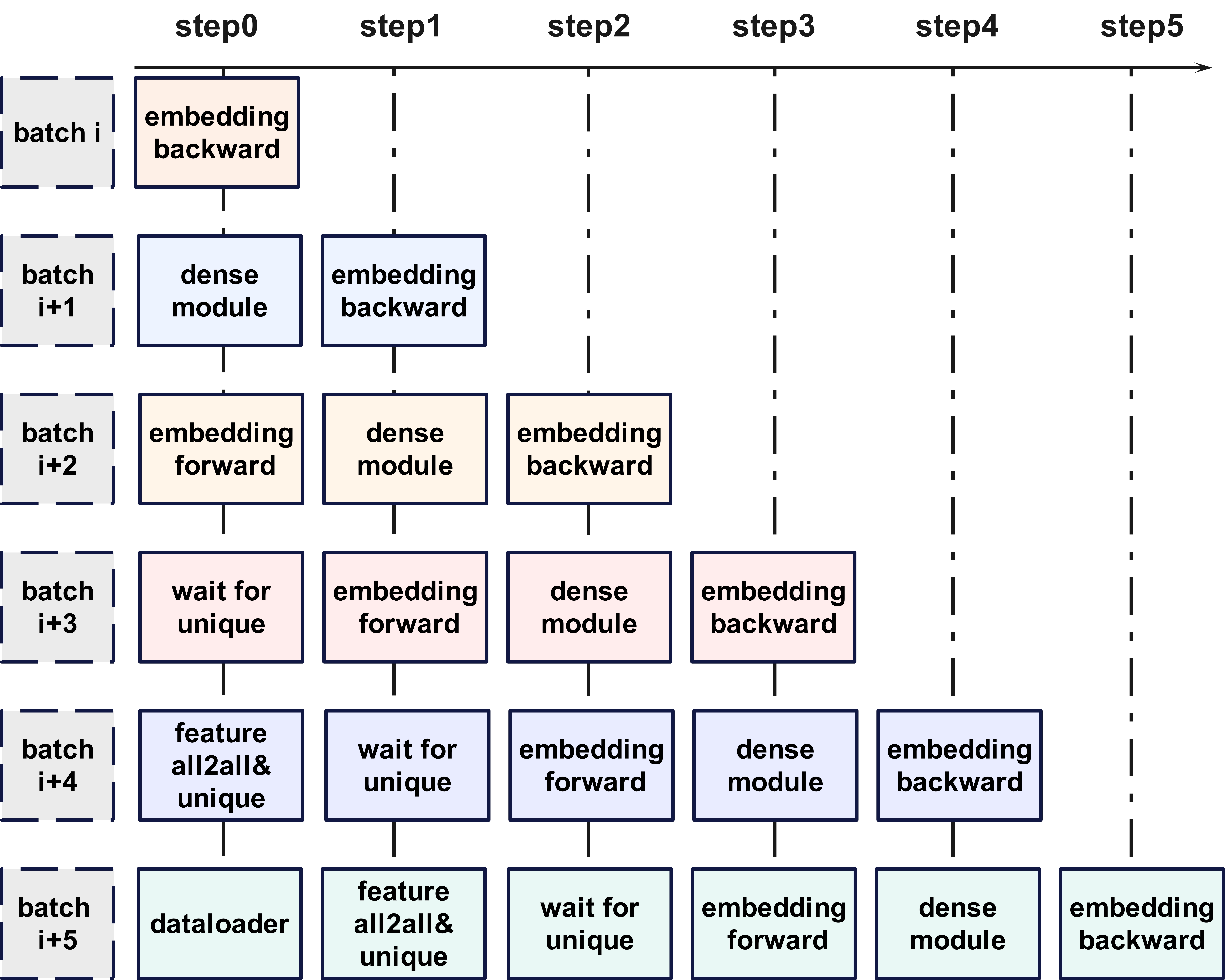}
    \caption{Fine-grained pipeline orchestration.}
    \label{fig:pipeline scheme}
\end{figure}

To maximize overall NPU utilization, we design a novel six-batch fine-grained pipeline that integrates the optimizations in Sections~\ref{sec:3.2.1} and~\ref{sec:3.2.2}. The key idea is to decouple the rigid CPU-NPU synchronization dependencies and overlap CPU and NPU operations, thereby sustaining NPU computation and eliminating idle periods. Compared to the implementation in~\cite{2022TorchRec}, we increase the pipeline depth to interleave NPU compute dispatch with CPU-side data processing and NPU communication, thereby maintaining continuous NPU saturation. In our design, the training step is delineated into the 6 stages as depicted in Figure~\ref{fig:pipeline scheme}: dataloader -> feature all-to-all \& CPU unique -> wait for unique -> embedding forward -> dense module (i.e., forward and backward) -> embedding backward.
The execution of this pipeline is sketched in Algorithm~1 in Appendix~\ref{sec:pipeline_algo}.

\begin{table}
    \centering
    \small 
    \caption{Communication latency analysis with the fine-grained pipeline orchestration.}
    \begin{tabular}{@{}lcccccccc@{}}
        \toprule
        \multirow{3}{*}{\textbf{Method}} & \multicolumn{2}{c}{\textbf{Computing}} & \multicolumn{2}{c}{\textbf{Communication}} & \multicolumn{2}{c}{\textbf{Comm. (Not Overlapped)}} & \multicolumn{2}{c}{\textbf{Free}} \\
        \cmidrule(lr){2-3} \cmidrule(lr){4-5} \cmidrule(lr){6-7} \cmidrule(lr){8-9}
        & Latency & Ratio & Latency & Ratio & Latency & Ratio & Latency & Ratio \\
        & (ms) & (\%) & (ms) & (\%) & (ms) & (\%) & (ms) & (\%) \\
        \midrule
        FuXi-large & 656.39 & 94.25 & 326.56 & 46.89 & 38.78 & 5.57 & 1.27 & 0.18 \\
        FuXi-long  & 1712.00 & 94.29 & 436.71 & 24.04 & 97.88  & 5.39 & 5.91 & 0.33 \\
        \bottomrule
    \end{tabular}
    \label{tab:fuxi-performance}
\end{table}

We conduct an empirical analysis on FuXi-large and FuXi-long in Table~\ref{tab:fuxi-performance}. With the proposed fine-grained pipeline orchestration, NPU idle time is reduced to less than 1\%, while non-overlapped communication remains less than 6\%. As a result, NPU computation accounts for 94\% of the end-to-end time. Taken together, the results provide strong evidence that the proposed optimizations are a good fit for GR training.

\subsection{Negative Sampling Optimization}
\label{sec:3.3}

In GR training for recall tasks, each positive sample is paired with a fixed number of independently sampled negatives, forming a contrastive loss that discriminates positives from negatives. However, scaling up negative sampling creates severe system bottlenecks dominated by embedding-related costs: HBM usage can exceed 20 GB, and loss computation contributes over 30\% of end-to-end latency. Moreover, the method requires storing large negative-sample embedding representations for every training instance, causing HBM consumption to grow linearly with the number of negatives and severely limiting scalability under memory constraints. To address this challenge, in this section, we propose a set of comprehensive negative-sampling optimization strategies that enable efficient recall training, including offloading, jaggedness-aware quantization, and 
reordering strategies for negative samples.

\subsubsection{Asynchronous Offloading for Negative Sampling}\

\begin{figure}
    \centering
    \includegraphics[width=0.7\linewidth]{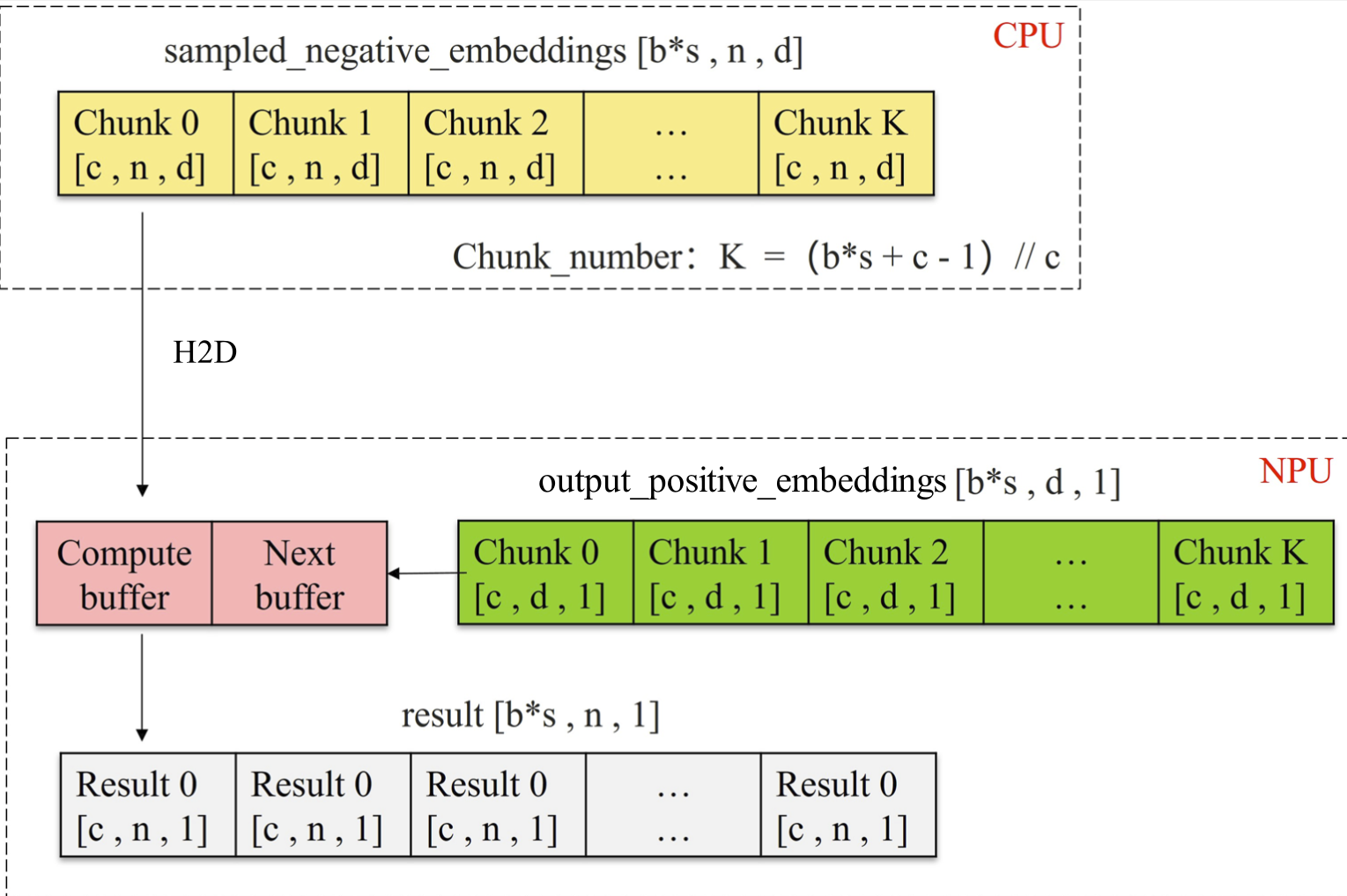}
    \caption{Asynchronous offloading for negative sampling.}
    \label{fig:NegativeSampleOffloading}
\end{figure}

In negative-sampling loss computation for GR models, the negative embedding tensor can occupy multiple gigabytes on a single NPU, making it the dominant memory consumer. Moreover, it remains resident in memory after the dot-product stage despite being unused elsewhere. As shown in Figure~\ref{fig:NegativeSampleOffloading}, we exploit the segment-wise independence of dot-product logit computation and introduce a ``CPU offloading + segmented fetching'' optimization. 
The logit computation at each valid position depends only on its local slice of the negative-embedding tensor, with no cross-segment dependencies. Moreover, the negative embeddings form a regular 3D dense tensor, so we can segment along the valid-position dimension and later concatenate the partial logits without extra bookkeeping. Based on this property (Figure~\ref{fig:NegativeSampleOffloading}), we offload negative embeddings from NPU HBM to CPU memory and fetch them back to the NPU segment by segment at a fixed granularity (e.g., 100 valid positions per segment). On the NPU, we use a double-buffered design (compute buffer + prefetch buffer): while the current segment is being processed, the next segment is asynchronously transferred into the prefetch buffer and then swapped in once the dot product finishes. After all segments are processed, we concatenate the per-segment logits and feed them into the original loss pipeline, thereby eliminating the HBM footprint previously occupied by the full negative-embedding tensor.

We investigate the performance of offloading optimization in Table \ref{tab:NegativeSampleOffloading_result} using the FuXi-large model. The results show that negative sampling offloading optimization reduces HBM usage by 7.3\% with 32 negative samples, 12.51\% with 64 samples, and 24.59\% with 128 samples. Notably, the memory savings grow superlinearly with the number of negative samples. As a result, despite the additional host–device transfer overhead, offloading remains a key enabler for recall training with large negative sets by substantially relaxing HBM constraints at scale.

\begin{table*}[t]
\centering
\small
\caption{HBM usage of FuXi-large with different numbers of negative samples.}
\begin{tabular}{cccccccc}
\toprule
\textbf{\# Negatives} & \textbf{Strategy} & \textbf{HBM Usage (GB)} &
\textbf{HBM Ratio (\%)}  \\
\midrule
\multirow{2}*{32}  & Baseline   & 22.21 & 34.70   \\
                   & Offloading & 17.42 & 27.22   \\
\midrule
\multirow{2}*{64}  & Baseline   & 31.64 & 49.44   \\
                   & Offloading & 23.44 & 36.63   \\
\midrule
\multirow{2}*{128} & Baseline   & 50.39 & 78.73   \\
                   & Offloading & 34.27 & 53.55   \\
\bottomrule
\end{tabular}
\label{tab:NegativeSampleOffloading_result}
\end{table*}

\subsubsection{Jaggedness-aware Quantization}\

\begin{figure}
    \centering
    \includegraphics[width=0.9\linewidth]{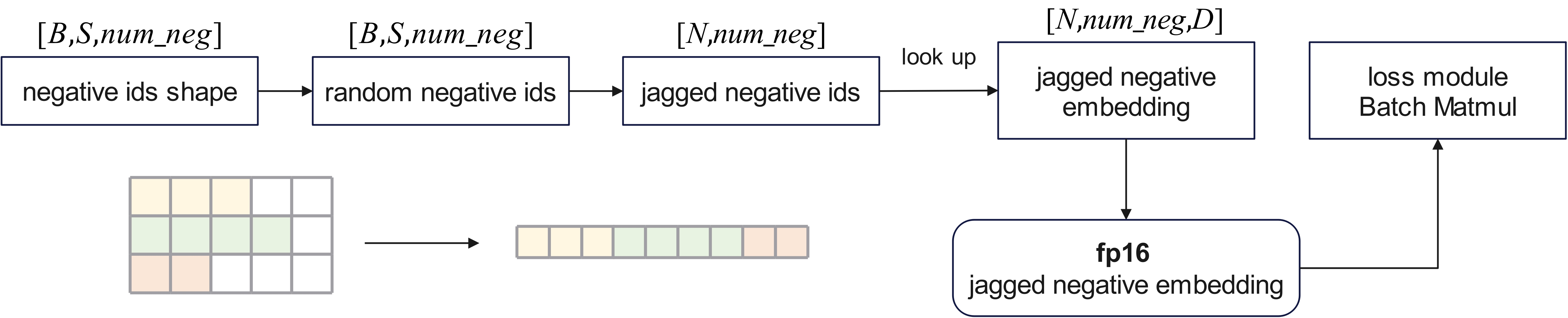}
    \caption{Jaggedness-aware quantization for negative sampling.}
    \label{fig:NegativeSampleQuantization}
\end{figure}

Owing to the jaggedness (i.e., variable-length sequences) discussed in Section~\ref{sec:3.1}, there exist significant variations in the number of negative samples across different users, which consequently leads to an irregular length structure of negative sample sequences. To address this issue, we adopt a jagged-tensor formulation for negative sampling, as illustrated in Figure~\ref{fig:NegativeSampleQuantization}. During data loading, we first sample negative IDs for the current batch from the global item ID pool according to the batch size, maximum sequence length, and the predefined number of negatives, forming dense ID matrices. We then apply the jagged structure to filter out IDs associated with semantically invalid (padded) positions, and construct a jagged ID tensor for embedding lookup. By removing invalid entries while preserving the original sequence structure, this approach reduces both storage and computation overhead for negative sampling. As a result, it lowers the memory footprint and improves scalability for large-scale sequence models.

To further reduce memory, we additionally apply quantization to negative sampling by storing and fetching negative-sample embeddings in half precision (FP16). In TorchRec, positive and negative samples are modeled as separate features but share the same embedding table; we therefore integrate an FP16 lookup path that is selectively triggered for negative samples, returning FP16 embedding vectors while leaving the rest of the pipeline unchanged.

\begin{figure}
    \centering
    \includegraphics[width=\linewidth]{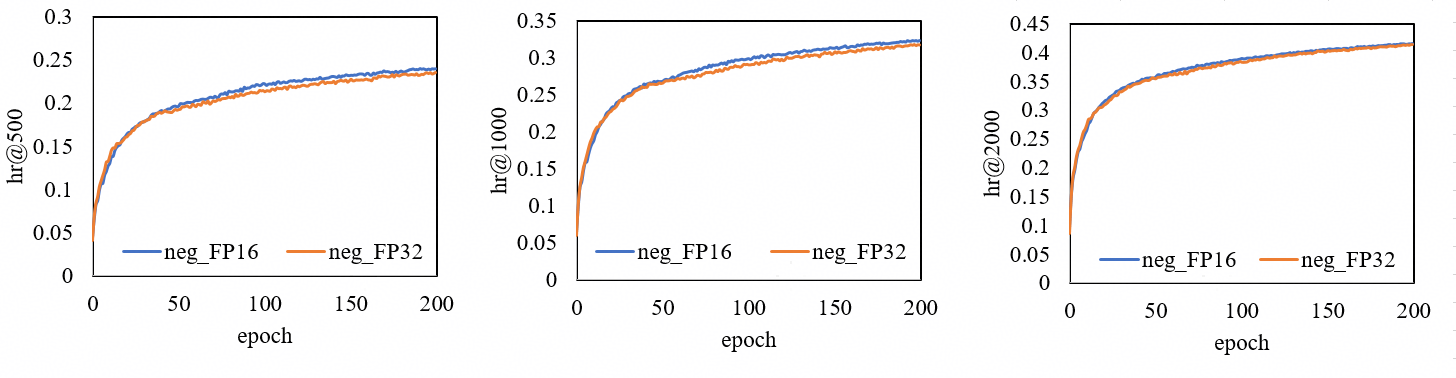}
    \caption{Accuracy after FP16 quantization for negative-sample embeddings.}
    \label{fig:NegativeSampleQuantization_result}
\end{figure}

As shown in Figure~\ref{fig:NegativeSampleQuantization_result}, the model shows only marginal differences from the baseline: 0.01\% in HR@2000 and 0.05\% in HR@1000. The measurement noise is small, indicating that the accuracy impact of FP16 negative embeddings is negligible. This is expected because FP16 retains 10 mantissa bits, and the quantization error is further attenuated by normalization and the fact that the logits are produced by a single matrix multiplication. In conclusion, FP16 quantization for negative-sample embeddings offers substantial memory savings with negligible impact on recommendation quality, making it a practical complement to jaggedness-aware negative sampling at scale.

\subsubsection{Information-Enhanced Negative Sampling Strategy via Logit Sharing}\

\begin{figure}
    \centering
    \includegraphics[width=0.7\linewidth]{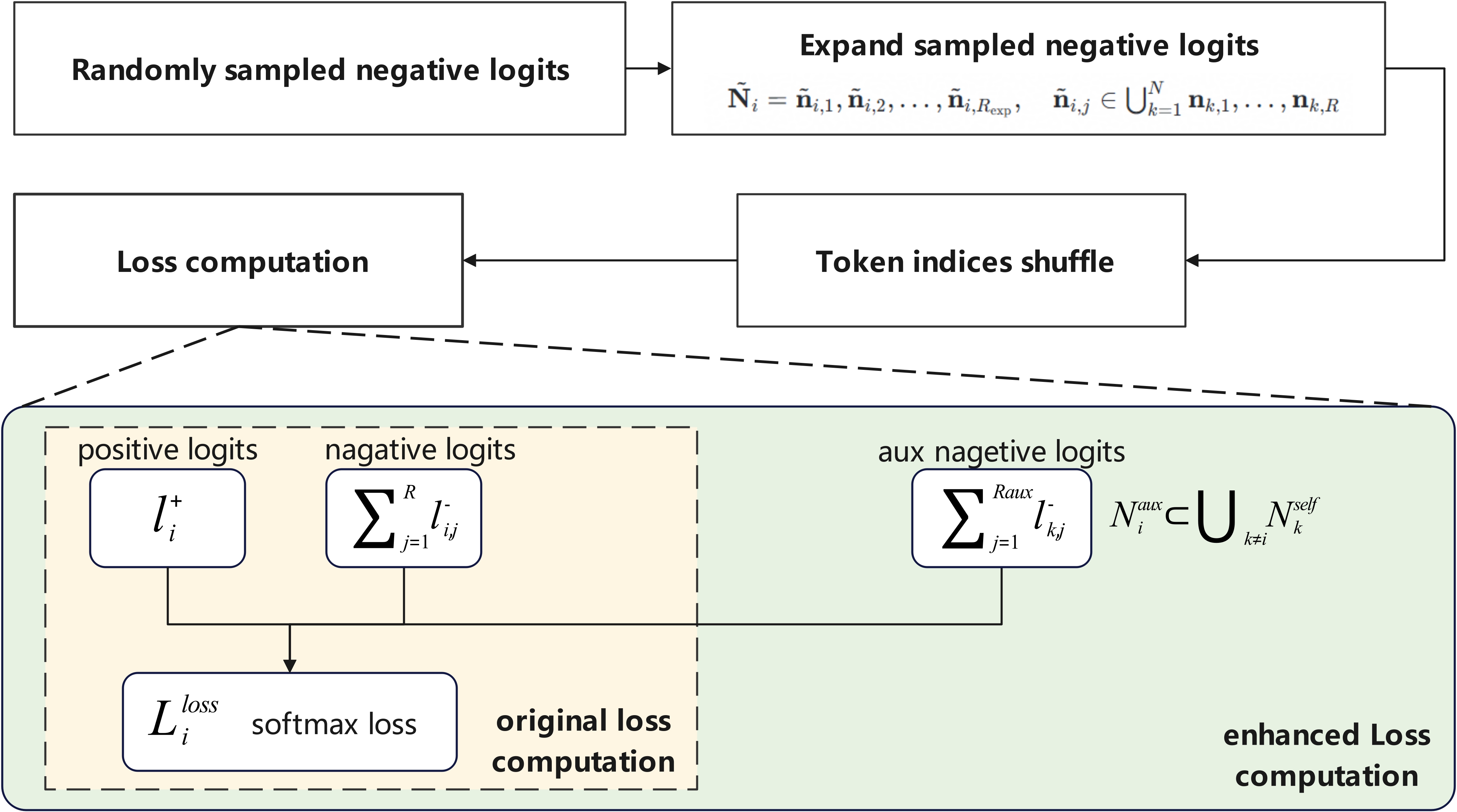}
    \caption{Intra-batch logit sharing with token-level shuffling.}
    \label{fig:NegativeSampleShared}
\end{figure}

To further alleviate excessive NPU memory consumption, redundant negative-sample information, and limited training efficiency under large-scale negative sampling, we propose an information-enhanced negative sampling strategy (Figure~\ref{fig:NegativeSampleShared}). The key idea is to exploit intra-batch sharing: for each target token, we reuse negative sample logits computed for other tokens in the same batch as auxiliary negatives. The negative sample logits represent the similarity scores derived from a batch matrix multiplication between the negative sample embeddings and the model output embeddings, and we concatenate them to expand the effective negative space without looking up additional negative embeddings. Meanwhile, to reduce the redundancy introduced by a fixed concatenation order, we apply a token-level shuffle to the extended negative logit set for each token. This randomizes the reconstructed set and mitigates token-level redundancy in the expanded negatives.

To incorporate the enhanced auxiliary negatives into training, we feed the target token's positive logit together with the reordered, expanded negative logits into a softmax-based contrastive loss. Let $N$ denote the total number of tokens and $R$ the number of original negative samples. For each token $i$, its negative set comprises its own negatives $N_i^{\text{self}}$ and an auxiliary subset drawn from other tokens, $N_i^{\text{aux}} \subseteq \bigcup_{k\neq i} N_k^{\text{self}}$.

The final loss optimization function for each token $i$ after auxiliary information enhancement is as follows: 
\begin{equation}
    Loss = -\log\left(\frac{\exp(l_i^+)}{\exp(l_i^+) + \sum_{j=1}^{R} \exp(\frac{\mathbf{o}_i^\top \cdot \mathbf{n}_{i,j}}{\tau}) + \Delta}\right)
\end{equation}
where $\Delta = \sum_{j=1}^{R_{\text{aux}}} \exp\left(\frac{\mathbf{o}_i^\top \cdot \mathbf{n}_{k,j}}{\tau}\right)$ is the auxiliary negative sample enhancement logits, $k \in R_{\text{aux}}$ and $k \neq i$. $l_i^+ = \frac{\mathbf{o}_i^\top \cdot \mathbf{p}_{i,j}}{\tau}$ and $l_i^- = \frac{\mathbf{o}_i^\top \cdot \mathbf{n}_{i,j}}{\tau}$ represent positive sample logits and original negative sample logits. This loss formulation enables the target sample to simultaneously receive discriminative constraints from its native negatives and integrate cross-sample discriminative signals from auxiliary negatives, thereby strengthening global candidate space discrimination capability.

We evaluate the negative sampling strategy in Table \ref{tab:NegativeSampleShared}. For compact models (e.g., tiny, small, and medium variants), expanding negative sampling logits by a factor of 2 yields accuracy fluctuations within a negligible range relative to the baseline, constituting statistically reasonable error and indicating accuracy parity. Conversely, for the FuXi-large model, logit expansion by a factor of 4 is required to achieve comparable accuracy to the baseline. When the model embedding dimension exceeds 1024 (i.e., the large variant), the batch size necessitates proportional reduction, resulting in insufficient data volume within the negative sampling set. Meanwhile, high-dimensional embeddings induce information redundancy among logits from numerous negative samples, causing the actual effective sample expansion dimension to fall below the configured k value. To ensure negative sampling diversity, a higher degree of stochastic expansion dimensionality must therefore be introduced. Based on empirical evaluation, using a sequence length of 2k and 128 negative samples as an example, we recommend the minimum expansion factor $k$ values for different embedding dimensions as shown in Table~\ref{tab:NegativeSampleShared_results}.

\begin{table*}[t]
\centering
\small
\caption{Recall performance of FuXi with different negative sampling strategies. $k$ denotes the expansion factor.}
\begin{tabular}{cccccccc}
\toprule
\textbf{Model} & \textbf{\# Negatives} & \textbf{HR@100} &
\textbf{HR@1000} &\textbf{NDCG@10} & \textbf{NDCG@200} & \\
\midrule
\multirow{2}*{FuXi-tiny} & 128 (baseline)     & 0.0633 & 0.2250 & 0.0059 & 0.0200  \\
                   & 64->128 ($k$=2) & 0.0619 & 0.2253 & 0.0061 & 0.0201  \\
\midrule
\multirow{2}*{FuXi-small} & 128 (baseline)     & 0.1020 & 0.3174 & 0.0114 & 0.0331  \\
                   & 64->128 ($k$=2) & 0.1060 & 0.3172 & 0.0128 & 0.0349   \\
\midrule
\multirow{2}*{FuXi-medium} & 128 (baseline)     & 0.1043 & 0.3148 & 0.0129 & 0.0343   \\
                   & 64->128 ($k$=2)& 0.1059 & 0.3135 & 0.0125 & 0.0346   \\
\midrule
\multirow{3}*{FuXi-large} & 128 (baseline)   & 0.0235 & 0.3204 & 0.0123 & 0.0347   \\
                   & 64->128 ($k$=2) & 0.0238 & 0.3136 & 0.0123 & 0.0341   \\
                   & 64->256  ($k$=4) & 0.0257 & 0.3183 & 0.0134 & 0.0360   \\
\bottomrule
\end{tabular}
\label{tab:NegativeSampleShared}
\end{table*}

\begin{table*}[]
\centering
\small
\caption{Minimum expansion factor $k$ values for different models.}
\begin{tabular}{cccccc}
\toprule
\textbf{Seq. Lengths} & \textbf{Total Logits} & \textbf{Model} & \textbf{Batch Size}& \textbf{\# Negatives}& \textbf{k} \\
\midrule
\multirow{4}*{2048}& \multirow{4}*{128}& FuXi-tiny &15 & 64 & \textbf{2}   \\
                   &                   & FuXi-small &8  & 64 & \textbf{2}   \\
                   &                   & FuXi-medium &5  & 64 & \textbf{2}   \\
                   &                   & FuXi-large&2  & 64 & \textbf{4}    \\
\bottomrule
\end{tabular}
\label{tab:NegativeSampleShared_results}
\end{table*}

\section{Conclusion}
\label{sec:conclusion}

We have presented \model, an Ascend-affinity training system for generative recommendation that systematically addresses the three fundamental bottlenecks arising when deploying GR models on NPU clusters: jagged computation redundancy, sparse-dense communication overhead, and excessive memory consumption from negative sampling. Through Ascend-affinity jagged fusion operators and dynamic load balancing, we reduce padding-induced redundancy and inter-device imbalance; through hierarchical sparse parallelism, semi-asynchronous training, and fine-grained pipeline orchestration, we achieve 94\% NPU utilization with near-linear scalability; and through asynchronous offloading, jaggedness-aware quantization, and intra-batch logit sharing, we enable large-scale recall training under tight HBM constraints. Evaluated on the KuaiRand-27K dataset, \model achieves 54.71\% MFU at 0.2B parameters with a cluster linearity of 0.97.

Looking ahead, several directions remain open. First, extending \model to support mixture-of-experts (MoE) architectures would further increase model capacity while introducing new challenges in expert-level load balancing and communication. Second, exploring sparse attention mechanisms tailored for Ascend NPUs could unlock training with even longer sequences (e.g., 16K--32K). Third, integrating automatic parallelism search to jointly optimize over the space of hybrid parallelism strategies would reduce manual tuning effort and improve portability across different cluster configurations. We hope that \model serves as a foundation for future research on efficient GR training for emerging AI accelerators.

\section*{Acknowledgments}
This work was supported in part by......

\bibliographystyle{unsrt}  
\bibliography{references}  

\clearpage
\appendix
\section{Experimental Settings}
\label{appendix:A}
We here introduce the experimental settings. All the experiments are conducted on 32-128 Ascend 910B1 NPUs across 4-16 NPU nodes with Kunpeng-920 ARM CPUs.

\paragraph{Dataset}
We use KuaiRand-27K, the largest and most widely adopted subset of the KuaiRand short-video recommendation dataset. It contains interaction logs from over 27,000 users with millions of items and tens of millions of user–video interactions collected over a one-month period. Each impression records rich multi-signal feedback, including watch time, click, like, follow, and completion rate. Compared with other KuaiRand variants, the 27K version provides significantly longer user–item interaction histories, resulting in much longer input sequences per user, as well as a larger and more realistic item space. These characteristics make KuaiRand-27K particularly suitable for evaluating model scalability and robustness under long-sequence scenarios.

\paragraph{Dataset Preprocessing}
To prepare KuaiRand-27K for the retrieval task, we remove negative interactions in which (1) the user explicitly indicates a dislike for the video, or (2) the user has no positive interactions, defined as at least one of the following behaviors: click, like, follow, comment, forward, or long-duration view.
We then apply 5-core filtering to ensure that each user has at least five interactions and each item has been interacted with by at least five users.
Subsequently, we group interactions by user and sort each user sequence chronologically. We adopt the leave-one-out strategy for train-test split, using all but the last item in each sequence for training and reserving the last item as ground truth for evaluating the model's retrieval performance.

\paragraph{Parameter Settings}
We evaluate several models, including SASRec, HSTU, and FuXi, by constructing four scaled variants (tiny, small, medium, and large) for each model to study their scalability. Specifically, the variants use item embedding dimensions of 128/256/512/1024 with a fixed sequence length of 2000. The corresponding backbones consist of 2/4/8/16 stacked blocks, each equipped with 8 attention heads and 16/32/64/128-dimensional query, key, and value projections.
To further validate the capability of handling ultra-long sequences, we introduce HSTU-long and FuXi-long variants with a sequence length of 4096, with all other settings aligned to their respective “large” variants.
We incorporate RAB by explicitly modeling both temporal and positional information: temporal dynamics are captured via bucketized time encodings (32 buckets) in HSTU and functional time encodings \cite{yi2025fuxigamma} in FuXi, while relative positional encodings are used to encode the order of interaction sequence.

The model is optimized using AdamW with a learning rate of $4\times10^{-3}$ and no weight decay. We apply dropout ($p=0.5$) to linear layers, adopt sampled softmax with 128 negative samples, and train with TF32 and asynchronous embedding updates to further improve computational efficiency.

\section{Details of Jagged Fusion Operators for Attention and RAB}
\label{appendix:rab}

We provide implementation details of jagged fusion operators for Attention and RAB, as follows.

\paragraph{Eliminating Unnecessary Conversions} Mismatches between the jagged format and the model’s internal representation constitute a major performance bottleneck for the operator. Unifying these formats removes costly dense–jagged conversions, reducing overhead and enabling further performance optimizations.

\paragraph{Tiling Strategies and Asynchronous Execution} 
Ascend NPUs process data in tiled chunks to balance compute load and efficiently utilize cache and memory resources. The choice of tiling strategy depends on data shapes, with rectangular tiling favored for large tensors and square tiling for smaller ones. Additionally, Ascend NPUs have low-level capabilities for efficient mass data copying, and copying does not block computing operations.

\paragraph{Load Balancing Across Device Computing Elements} 
The efficiency of the RTB backward pass depends on balanced utilization of scalar and vector computing elements within a device. We offload regular, data-parallel computations to vector units and reserve scalar units for irregular operations, requiring only lightweight data packing. This design significantly improves parallelism and is especially effective for the RTB backward pass, where naive implementations heavily overload scalar units.

\section{Convergence Analysis of Semi-Async Training Strategy}
\label{sec:convergence}
To analyze the convergence of the strategy, we introduce sparsity and delay steps to characterize the recommendation training system. The sparsity of samples in recommendation systems is an inherent property, denoted by $\alpha$, indicating the probability of collision between samples of different steps. $N_{emb}$ is the total number of unique features. If every feature appears with equal probability, $\alpha$ is $\frac{1}{N_{emb}}$. If a single feature appears in every sample, $\alpha$ is maximal (1), indicating no sparsity for that feature. The delay steps of sparse module asynchrony, denoted by $\tau$, represent the number of steps between forward computation and backward updates for embedding weights. The convergence of the semi‑asynchronous system satisfies:

\begin{equation}
    \frac{1}{T}\sum_{t=0}^{T-1} \mathbb{E}[\|\nabla f(w_t)\|^2] \leq O\left(\frac{\sqrt{L}\sigma}{\sqrt{T}} + \frac{L}{T} + \frac{\alpha L \tau}{T}\right),
\end{equation}
where the left‑hand side denotes the average expected squared gradient norm over $T$ iterations, which serves as a standard measure of stationarity in non‑convex optimization. 

On the right‑hand side, $L$ is the Lipschitz constant of the gradient, $\sigma$ bounds the variance of the stochastic gradients. Note that the first two terms are exactly the convergence rate of vanilla SGD. And $O(\alpha L \tau/T)$ is the delay penalty induced by the asynchronous embedding updates. 

When $\alpha \ll 1$ (extremely sparse), the third term becomes negligible. Even with a large $\tau$, as long as $\alpha$ is sufficiently small, the convergence remains close to synchronous counterpart. This demonstrates that features with sparse updates and a low probability of collision between different steps can alleviate the impact of delayed updates on convergence. 

A similar convergence analysis can be derived for optimizers like Adagrad. In our system, $\alpha \ll 1$ and $\tau = 1$, making the delay penalty negligible in theory. Empirical results confirm that the accuracy of this semi-asynchronous strategy nearly matches that of fully synchronous training, as shown in Table~\ref{tab:Semi-Async training comm} of the main text, with a maximum accuracy decrease of 0.26\% relative to the baseline after 200 epochs. This semi-asynchronous training scheme enables misalignment between sparse and dense batches, allowing:
\begin{enumerate}
    \item The backward pass of batch $i$'s dense model to overlap with the forward pass of batch (i+1)'s sparse model.  
    \item The forward pass of batch $(i+2)$'s dense model to overlap with the backward pass of batch $(i+1)$'s sparse model.  
\end{enumerate}

\section{6-Batch Pipelined Overlapping Execution}
\label{sec:pipeline_algo}

In this appendix we present the algorithm of the fine-grained pipeline orchestration. Note that throughout training, cross-device transfers (host-to-device and device-to-host) introduce mandatory synchronization points. To avoid pipeline stalls, we implement these transfers asynchronously, enabling non-blocking instruction dispatch from the CPU to the NPU.

\begin{center}
\label{algorithm:pipeline}
    \includegraphics[width=1.0\linewidth]{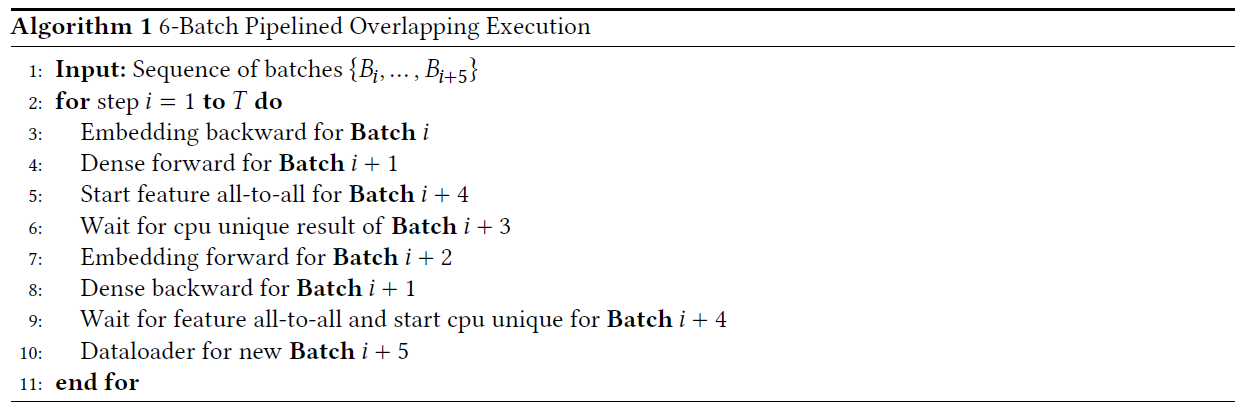}
\end{center}

\end{document}